\documentclass[12pt]{iopart}
\usepackage{iopams}

\input epsf.sty

\newlength{\TZ}
\newcommand{\BEQ}{\begin{equation}}     
\newcommand{\BEA}{\begin{eqnarray}}
\newcommand{\BD}{\begin{displaymath}}
\newcommand{\EEQ}{\end{equation}}       
\newcommand{\EEA}{\end{eqnarray}}
\newcommand{\ED}{\end{displaymath}}
\newcommand{\eps}{\varepsilon}          
\newcommand{\D}{{\rm d}}                
\newcommand{\II}{{\rm i}}               
\newcommand{\wit}[1]{\widetilde{#1}}    
\newcommand{\wht}[1]{\widehat{#1}}      
\newcommand{\rar}{\rightarrow}          

\renewcommand{\vec}[1]{\boldsymbol{#1}} 

\newcommand{\annexe}[1]{\setcounter{equation}{0}\setcounter{subsection}{0}
\section*{Appendix. #1}
\renewcommand{\theequation}{A\arabic{equation}}
              \renewcommand{\thesection}{A} }

\catcode`\@=11
\def\numberbysection{\@addtoreset{equation}{section}
        \def\theequation{\thesection.\arabic{equation}}}

\begin{document}

\title[Quasiprimary operators in local scale-invariance]{On the identification 
of quasiprimary scaling operators in local scale-invariance}

\author{Malte Henkel$^{a,b,c}$, Tilman Enss$^d$ and Michel Pleimling$^{e,f}$} 
\address{$^a$Laboratoire de Physique des 
Mat\'eriaux,\footnote{Laboratoire associ\'e au CNRS UMR 7556} 
Universit\'e Henri Poincar\'e Nancy I, \\ 
B.P. 239, F -- 54506 Vand{\oe}uvre l\`es Nancy Cedex, 
France\footnote{permanent
address}}
\address{$^b$Isaac Newton Institute of Mathematical Sciences, 
20 Clarkson Road,\\ Cambridge CB3 0EH, England}
\address{$^c$Dipartamento di Fisica/INFN - Sezione de Firenze, 
Universit\`a di Firenze,\\ I -- 50019 Sesto Fiorentino, Italy}
\address{$^d$INFM-SMC-CNR and Dipartamento di Fisica, 
Universit\`a di Roma ``La Sapienza'', Piazzale A. Moro 2, 
I -- 00185 Roma, Italy}
\address{$^e$Institut f\"ur Theoretische Physik I, 
Universit\"at Erlangen-N\"urnberg, \\
Staudtstra{\ss}e 7B3, D -- 91058 Erlangen, Germany}
\address{$^f$ Department of Physics, Virginia Polytechnic Institute and State University,\\
Blacksburg, VA 24061-0435, USA}

\begin{abstract}
The relationship between physical observables defined in lattice models
and the associated \mbox{(qua\-si-)}primary scaling operators of
the underlying field-theory is revisited. In the context of local
scale-invariance, we argue that this relationship is only defined up to
a time-dependent amplitude and derive the corresponding generalizations 
of predictions for two-time response and correlation functions.
Applications to non-equilibrium critical dynamics of several systems, with a 
fully disordered initial state and vanishing initial magnetization, 
including the Glauber-Ising model, the
Frederikson-Andersen model and the Ising spin glass are discussed. The critical 
contact process and the parity-conserving non-equilibrium kinetic Ising model 
are also considered. 
\end{abstract}

\submitto{\JPA }
\maketitle

\setcounter{footnote}{0}

The analysis of the collective behaviour of many-body systems is greatly
helped in situations where some scale-invariance 
allows an efficient description
through field-theoretical methods. A necessary requirement for the application
of these is the possibility to identify the physical observables typically
defined in terms of a lattice model, e.g. $\sigma_{\vec{r}}$ for the
order-parameter at the site $\vec{r}$, with a continuum field $\phi(\vec{r})$
(called a {\em scaling  operator} \cite{Card90}) 
with well-defined scaling properties 
$\phi(\vec{r}) = \mathfrak{b}^{-x} \phi(\vec{r}/\mathfrak{b})$.
In other words, one generally expects that the correspondence ($\mathfrak{a}$ 
is the lattice constant)
\BEQ \label{1}
\sigma_{\vec{r}} \rar  \mathfrak{a}^{-x} \phi(\vec{r})
\EEQ
can be defined in equilibrium systems or more generally steady-states of
non-equilibrium systems, see e.g. \cite{Drou88,Card90,Mont94}. In addition,
in equilibrium systems one expects the same sort of relationship to hold true
where $\phi(\vec{r})$ is now a primary scaling operator of a
conformal field-theory and allows
space-dependent rescaling factors 
$\mathfrak{b}=\mathfrak{b}(\vec{r})$ \cite{Card90}. 

In this letter, we reconsider this correspondence for systems with
dynamical scaling and far from equilibrium, as it occurs  for example in
ageing phenomena. Concrete examples are phase-ordering kinetics or
non-equilibrium critical dynamics, see \cite{Bray94a,Cala05,Henk06} for reviews.
Among the main quantities of interest are the two-time autocorrelation function
$C(t,s)$ and the autoresponse function $R(t,s)$
\BEA
C(t,s) &=& \Bigl\langle \phi(t,\vec{r}) \phi(s,\vec{r})\Bigr\rangle 
\:=\: s^{-b} f_C(t/s) 
\nonumber \\
R(t,s) &=& \left. \frac{\delta\langle \phi(t,\vec{r})\rangle}{\delta
h(s,\vec{r})}  \right|_{h=0} 
\:=\: \left\langle  \phi(t,\vec{r}) \wit{\phi}(s,\vec{r})\right\rangle
\:=\:  s^{-1-a} f_R(t/s)
\label{CR}
\EEA
where $\wit{\phi}$ is the response field in the Janssen-de Dominicis formalism
\cite{deDo78,Jans92}, $a$ and $b$ are ageing exponents and $f_C$ and $f_R$
are scaling functions such that $f_{C,R}(y)\sim y^{-\lambda_{C,R}/z}$ for 
$y\gg 1$. These scaling forms are only valid in the scaling
regime where $t,s\to\infty$ {\em and} $y=t/s>1$ fixed. Until recently, the scaling
(\ref{CR}) has only been studied for systems with a fully disordered initial state with
mean initial magnetization $m_0=\langle \phi(0,\vec{r})\rangle=0$. The study of the effects of a non-vanishing 
initial magnetization on the ageing behaviour is only beginning 
\cite{Anni06,Cala06,Fedo06}. We stress that in the kind of system under
consideration invariance under time-translations is broken. 
In an attempt to try to derive the form of the scaling
functions in a model-independent way it has been argued \cite{Henk02} that
the scaling operators $\phi$ and $\wit{\phi}$ should transform covariantly
under a larger group than mere dynamical scale-transformations. If such an
invariance exists, one may call it a {\em local scale-invariance (LSI)}.\footnote{All existing tests of LSI have been performed for $m_0=0$.} The 
infinitesimal generators of local scale-invariance read
\cite{Henk02,Pico04,Henk05a}
\BEQ
X_0 = - t \partial_t - \frac{x}{z} \;\; , \;\;
X_1 =  - t^2 \partial_t -\frac{2}{z} \left( x+\xi\right) t
\label{gl:X0X1}
\EEQ
where for simplicity we have suppressed the terms acting on the space 
coordinates which are not important for what follows. We  have also not
written down the further generators of LSI which do not  modify the
time $t$ but only act on the space coordinates $\vec{r}$, and the absence
of any scaling of $m_0$ means that we are restricting ourselves to the case $m_0=0$
throughout. Here $x$ is the
scaling dimension of the scaling operator 
$\phi(t,\vec{r})=\mathfrak{b}^{-x/z}\phi(t/\mathfrak{b}^z,\vec{r}/\mathfrak{b})$
where  $z$ is the dynamical exponent and $\xi$ is a constant. 
It is the purpose of this letter to clarify the meaning of this constant $\xi$. 

Motivated by the analogy with two-dimensional conformal invariance, we
generalize the dilatation generator $X_0$ and the generator $X_1$ of
`special' transformations as follows to all $n\geq 0$
\BEQ
X_n = - t^{n+1} \partial_t  - \frac{x}{z} (n+1) t^n - \frac{2\xi}{z} n t^n
\EEQ
such that the commutator $[X_n,  X_m]=(n-m) X_{n+m}$ holds for all
$n,m\in \mathbb{N}_0$ (with the convention $0\in \mathbb{N}_0$).\footnote{This
is the unique semi-infinite extension of the algebra 
$\langle X_0,X_1\rangle$ which does
not introduce further differential operators into $X_n$ and is compatible with 
eq.~(\ref{gl:X0X1}).} 
Next, the global form of these transformations reads as follows. If 
$t=\beta(t')$ such that $\beta(0)=0$, then $\phi(t)$ transforms as
\BEQ
\phi(t) = \dot{\beta}(t')^{-x/z} 
\left( \frac{t'  \dot{\beta}(t')}{\beta(t')}\right)^{-2\xi/z} 
\phi'(t')
\EEQ
where again the space-dependence of $\phi$ was suppressed. The infinitesimal 
generators $X_n$ are recovered for $\beta(t)=t+\epsilon t^{n+1}$, with
$|\epsilon|\ll 1$. From this, it is clear
that $\phi$ is not transforming as an usual primary scaling operator. But if
one defines $\Phi(t) := t^{-2\xi/z} \phi(t)$
the scaling operator $\Phi(t)$ becomes a conventional primary scaling operator
of LSI, viz.
\BEQ \label{gl:PhiTR}
\Phi(t) = \dot{\beta}(t')^{-(x+2\xi)/z}\;  \Phi'(t')
\EEQ
but with a modified scaling dimension $x\to x+2\xi$. 
In other words, if time-dependent observables of lattice models
$\sigma_{\vec{r}}(t)$ can be related to a primary scaling operator
$\Phi(t)$ at all, it should be via the relation
\BEQ \label{2}
\sigma_{\vec{r}}(t) \rar \mathfrak{a}^{-x}\: \phi(t) = \mathfrak{a}^{-x}\:  
t^{2\xi/z}\, \Phi(t)
\EEQ
rather than by eq.~(\ref{1}). Of course, (\ref{2}) is only possible because
of the absence of time-translation invariance. 
We emphasize that the scaling of $\phi$ is unusual in that under a dilatation
$t\to\mathfrak{b}^z t$ the scaling dimension remains $x$ but for more general
scale transformations a new effective scaling dimension $x+2\xi$ appears. 

As  a simple application, consider the two-time autoresponse function. 
For qua\-si\-pri\-ma\-ry scaling operators $\Phi(t)$ and $\wit{\Phi}(s)$ with 
scaling dimensions $x$ and $\wit{x}$, respectively, local scale-invariance
with $m_0=0$ predicts $\langle \Phi(t)\wit{\Phi}(s)\rangle = (t/s)^{(\wit{x}-x)/z} 
(t-s)^{-(x+\wit{x})/z}$, up to normalization, as shown in \cite{Henk02}. 
In view of (\ref{2}), the physical autoresponse function rather reads
\newpage \typeout{ *** hier ist ein Seitenvorschub von Hand *** }
\BEA
R(t,s) &=& \left\langle \phi(t) \wit{\phi}(s) \right\rangle 
= \left\langle t^{2\xi/z} \Phi(t) s^{2\wit{\xi}/z} \wit{\Phi}(s) \right\rangle
\nonumber \\
&=& s^{-(x+\wit{x})/z} \left(\frac{t}{s}\right)^{(2\wit{\xi}+\wit{x}-x)/z} 
\left(  \frac{t}{s}-1\right)^{-(x+\wit{x}+2\xi+2\wit{\xi})/z}
\nonumber \\
&=&  s^{-1-a} \left(\frac{t}{s}\right)^{1+a'-\lambda_R/z}  
\left( \frac{t}{s}-1\right)^{-1-a'}
\label{R}
\EEA
(up to normalization) 
and the effective scaling dimensions of $\Phi(t)$ and $\wit{\Phi}(s)$
are read off from eq.~(\ref{gl:PhiTR}) to be now $x+2\xi$ and
$\wit{x}+2\wit{\xi}$, respectively. In the last line, we have
reintroduced the standard exponents $a$, $a'$ and $\lambda_R$ and hence 
reproduce the result quoted in \cite{Henk05a}. Early discussions of local
scale-invariance had assumed $a'=a$ from the outset. 
In the appendix, we discuss the scaling form of the autocorrelator 
$C(t,s)$ in those cases where $z=2$.     
 
\begin{table}
\caption[Tabelle 1]{Values of the exponents $a$, $a'$ and $\lambda_R/z$ in 
several non-disordered and a few  glassy systems which are at a critical point
of their stationary state.  
If a numerical result is quoted
without an error bar it is taken form the literature, otherwise the 
numbers in brackets give our estimate of the uncertainty in the last digit(s). 
{\sc nekim} stands for non-equilibrium kinetic Ising model with conserved parity 
and {\sc fa} stands for the Frederikson-Andersen model. The methods of 
calculation of the two-time autoresponse are {\sc d}: direct space, 
{\sc  p}: momentum space, {\sc a}: alternating external field; {\sc e} refers
to an exact solution and {\sc n} to a numerical study. 
\label{Tabelle1}
}
\begin{center}
\begin{tabular}{||l|rrr|cl||} \hline\hline\hline
model & $a$ & $a'-a$  & $\lambda_R/z$ & Method & Ref.  \\\hline\hline
OJK-model & $(d-1)/2$ & $-1/2$  & $d/4$ & {\sc d,e} & \cite{Bert99,Maze04,Henk05a}\\\hline
$1D$ Ising & 0 & $-1/2$  & $1/2$ & {\sc d,e} & \cite{Godr00,Pico04}\\\hline
$2D$ Ising & $0.115$ & $-0.187(20)$ & $0.732(5)$ & {\sc p,n}& \cite{Plei05}\\\hline
$3D$ Ising & $0.506$ & $-0.022(5)$  & $1.36$ & {\sc p,n}& \cite{Plei05}\\\hline
$1D$ contact process & $-0.681$ & $+0.270(10)$ & $1.76(5)$ & {\sc d,n} & \cite{Enss04,Hinr06} \\\hline
$1D$ {\sc nekim}     & $-0.430$ & $-0.09$      & $0.56$    & {\sc d,n} & \cite{Odor06} \\\hline\hline
{\sc fa}, $d>2$ & $1+d/2$ & $-2$  & $2+d/2$ & {\sc p,e} & \cite{Maye06} \\
{\sc fa}, $d=1$ & $1$ & $-3/2$ & $2$ & {\sc p,e} & \cite{Maye06,Maye04}\\\hline
$3D$ Ising spin glass & $0.060(4)$ & $-0.76(3)$ & $0.38(2)$ & {\sc a,n} & \cite{Henk05a} \\ \hline\hline\hline 
\end{tabular}\end{center}
\end{table}

It appears that the more general correspondence (\ref{2}) and
consequently the response (\ref{R}) with $a'\ne a$ actually occurs
in non-equilibrium critical dynamics, as we shall now illustrate in a few
examples. We stress that in the models considered here (with the only
exception of the contact process) we always use a fully disordered initial
state with a vanishing initial magnetization $m_0=0$.\footnote{From LSI, it is then
easy to see that the time-dependent magnetization $m(t)=m_0=0$ for all times,
in agreement with the Monte Carlo and the exact results. On the other hand, if
initially $m_0>0$, one has the regime of short-time dynamics with 
$m(t)\sim t^{\theta}$ \cite{Jans89} before the long-time decay 
$m(t)\sim t^{-\beta/(\nu z)}$ \cite{Fish76}. The scaling of 
two-time observables has been recently discussed
in \cite{Anni06,Cala06,Fedo06} and it was shown that for $m_0\ne 0$ the
universal scaling behaviour is different from the one found for $m_0=0$. 
An extension of LSI to non-equilibrium critical dynamics 
with non-vanishing initial magnetizations is an open problem to which to hope to return elsewhere.}
In table~\ref{Tabelle1} we collect results on the exponents $a$, $a'$  and
$\lambda_R/z$ in some models with a critical stationary  state and 
where $a'\neq a$.\footnote{In table~\ref{Tabelle1}, {\sc d,e} 
means that the exact response agrees with (\ref{R}) 
with the given values of the
exponents, while {\sc p,e} means that there is exact agreement with 
(\ref{Rq}).} 
In several cases, these exponents can be read off from the
exact solution, i.e., for the magnetic response in the OJK-model
\cite{Bert99,Maze04} and the $1D$ Glauber-Ising model at 
zero temperature \cite{Godr00} 
or else the energy response in the zero-temperature Frederikson-Andersen 
model \cite{Maye06,Maye04}. 

\begin{figure}[t]
\centerline{\epsfxsize=72mm\epsffile{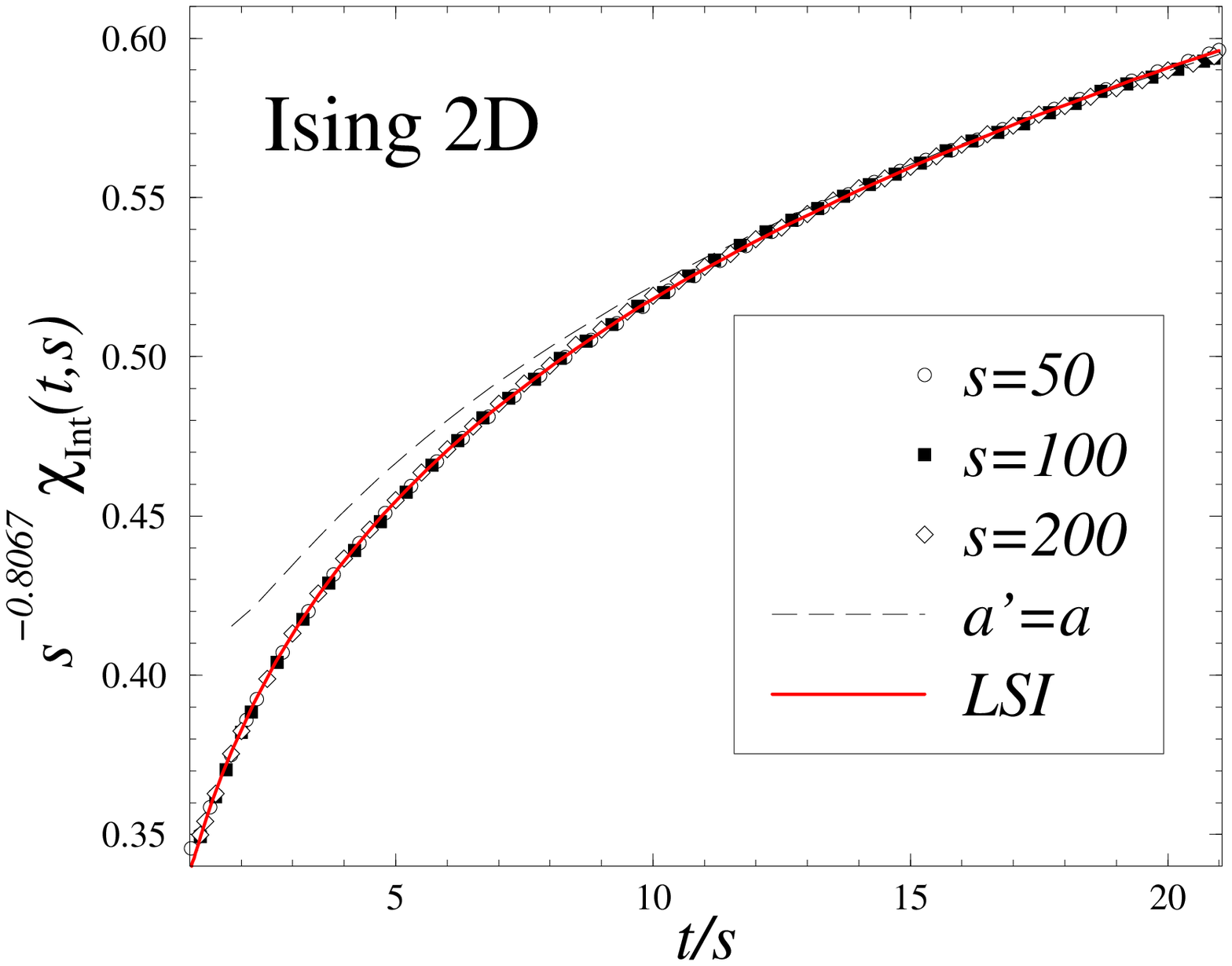} ~~~\epsfxsize=72mm
\epsffile{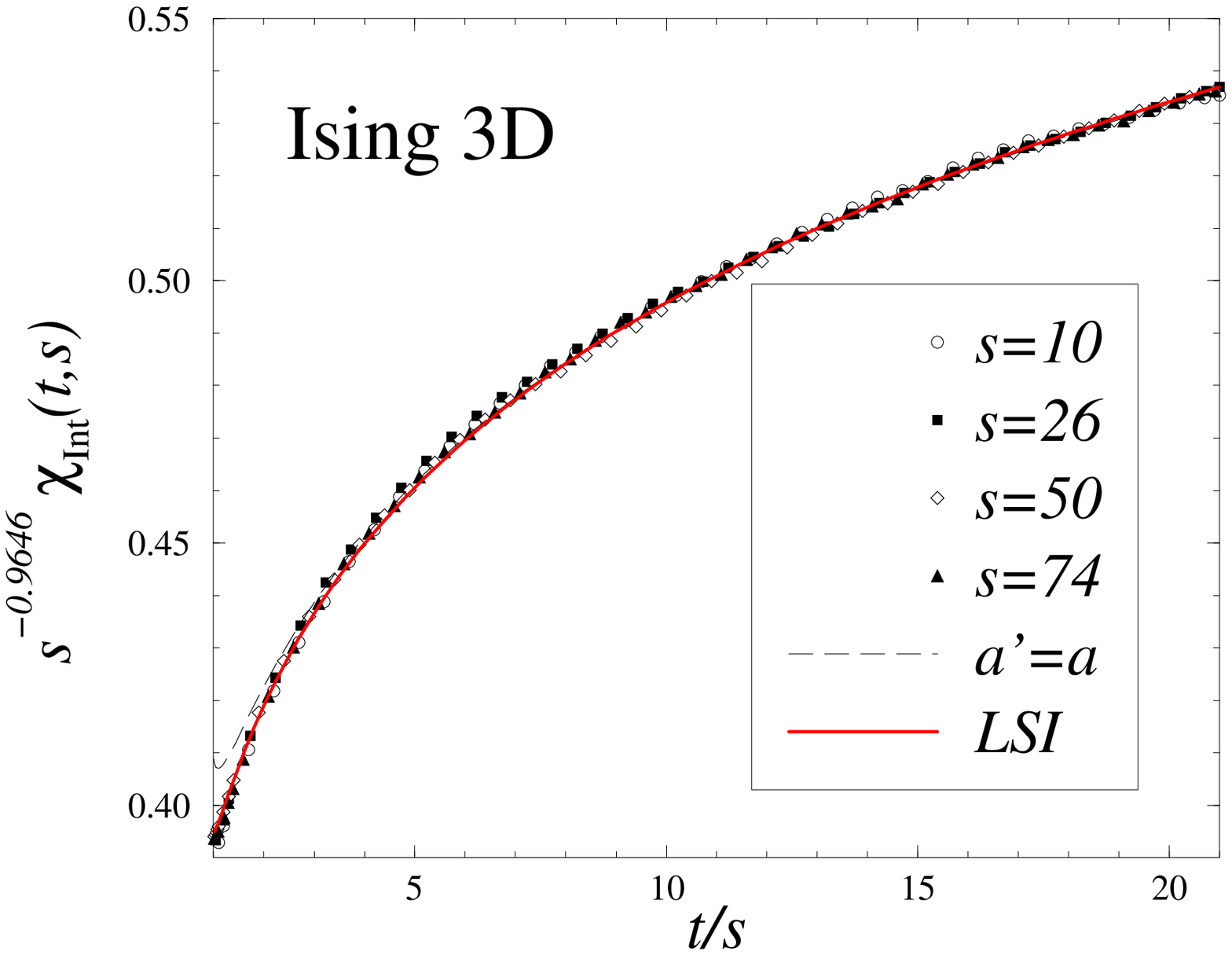}}
\caption[Abbildung 1]{Intermediate susceptibility $\chi_{\rm Int}(t,s)$ in
momentum space in the (a) $2D$ and (b) $3D$ critical Ising model, for
several values of the waiting time $s$. The full curve is the LSI prediction
eq.~{\protect (\ref{gl:IELchiint},\ref{gl:IELchiint2})}  
with the exponents as listed
in {\protect table~\ref{Tabelle1}}. The dashed line corresponds to the case 
$a'=a$. \label{Abb1}}
\end{figure}

Another interesting test case is provided  by the critical Ising model in $2D$
and $3D$. Indeed, it was pointed out some time ago that the numerical 
calculation of the two-time response function 
$\wht{R}_{\vec{q}}(t,s) = \int_{\mathbb{R}^d} \!\D  \vec{r}\: R(t,s;\vec{r})
e^{-\II \vec{q}\cdot\vec{r}}$ in {\em momentum} space provides
a more sensitive test on the form of its scaling function than in direct 
space \cite{Plei05}.
The methods of LSI can be readily adapted to momentum space and the
analogue of (\ref{R}) is, again up to normalization and for $m_0=0$
\BEQ \label{Rq}
\wht{R}_{\vec{0}}(t,s) = s^{-1-a+d/z}
\left(\frac{t}{s}\right)^{1+a'-\lambda_R/z}
\left(\frac{t}{s}-1\right)^{-1-a'+d/z}
\EEQ  
Since measurements of autoresponse functions are much affected by statistical
noise, one often rather studies integrated response functions. 
Here we consider
\BEQ \label{gl:IELchiint}
\chi_{\rm Int}(t,s) := \int_{s/2}^{s} \!\D u\: \wht{R}_{\vec{0}}(t,u) 
= \chi_0 s^{-a+d/z} f_{\chi}(t/s)
\EEQ
which is free from effects which mask the true scaling behaviour in several
other variants of integrated responses \cite{Plei05}.
The scaling function $f_{\chi}(y)$ follows from LSI, eq.~(\ref{Rq}): 
\BEA \label{gl:IELchiint2}
f_{\chi}(y) &=&  y^{(d-\lambda_R)/z} \left[ 
{}_2F_{1}\left(1+a'-\frac{d}{z},\frac{\lambda_R}{z}-a;1+\frac{\lambda_R}{z}-a;
\frac{1}{y}\right) \right.
\nonumber \\
& & \left. - 2^{a-\lambda_R/z} 
{}_2F_{1}\left(1+a'-\frac{d}{z},\frac{\lambda_R}{z}-a;1+\frac{\lambda_R}{z}-a;
\frac{1}{2y}\right) \right]
\EEA
and where $_2F_1$ is Gauss' hypergeometric function. In figure~\ref{Abb1} 
we compare  simulational data \cite{Plei05} with this prediction for both the
$2D$ and $3D$ critical Ising model with non-conserved heat-bath dynamics.
It had already been observed before \cite{Plei05} that local 
scale-invariance with the additional assumption $a'=a$ does not agree with
the numerical data in $2D$ and only marginally so in $3D$ and we confirm this
finding. However, we also see that the data can be perfectly matched by LSI,
within the numerical precision, if $a$ and $a'$ are allowed to be different. 
We did check that the integrated TRM response functions in direct space 
as studied in \cite{Henk01} do not change appreciably with $a'-a$. 

\begin{figure}
\centerline{\epsfxsize=85mm\epsffile{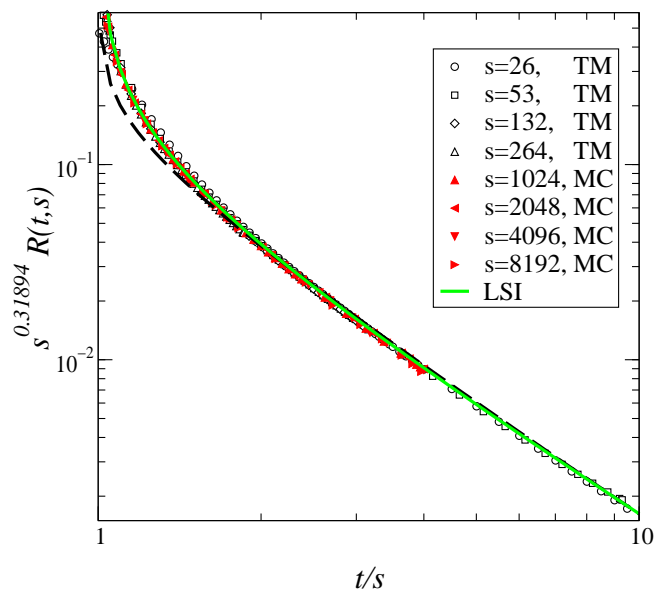}
}
\caption[Abbildung 2]{Autoresponse function for the critical $1D$
contact process for several waiting times $s$. The data labelled {\sc tm}
come from the transfer matrix renormalization group {\protect \cite{Enss04}} 
and {\sc mc} denotes Hinrichsen's Monte Carlo data 
{\protect \cite{Hinr06}}. 
The dashed line corresponds to the case $a'=a$ and the full curve
gives the LSI prediction {\protect eq.~(\ref{R})} with the exponents
as listed in {\protect table~\ref{Tabelle1}}. 
\label{Abb2}
}
\end{figure}

A similar conclusion can also be drawn for the $1D$ critical contact process. 
It has been shown recently that the phenomenology of  ageing can also
be found in critical stochastic processes  although these do {\em not} satisfy
detailed balance and have a non-equilibrium steady-state 
\cite{Oerd98,Rama04,Enss04}.
In figure~\ref{Abb2} we compare the numerical data obtained directly for
$R(t,s)$ either from the LCTMRG \cite{Enss04} or Monte Carlo simulations
\cite{Hinr06}. It is satisfying that the data from both methods are  
consistent with each other in the scaling regime, where $s$ and $t-s$ are
both large enough. Again, we observe an almost perfect agreement
with eq.~(\ref{R}), provided $a'\ne a$.\footnote{Hinrichsen quotes 
$\lambda_R/z\approx 1.75$ and $1+a'\approx0.59$ \cite{Hinr06} 
in good agreement with our estimates. The contact process is the only known 
example where $a'-a>0$.} 

One the other hand, when one looks closer at the region where
$t/s\lessapprox 1.1$, one does observe deviations of the data
from (\ref{R}) \cite{Hinr06}. In trying to analyze this, recall 
that non-equilibrium {\em critical} dynamics is special in the sense
that both the ageing regime (where $t-s\sim {\rm O}(s)$)
and the quasistationary regime (where $t-s\ll s$) display dynamical
scaling with the same length scale $L(t)\sim t^{1/z}$, where $z$ is the
equilibrium dynamical critical exponent. Hence one usually expects 
some `crossover' to occur. In terms of the response function, this
might be formalized by writing
$R={\cal R}(s/\tau_{\rm *},(t-s)/\tau_{\rm *},s)$ where  $\tau_*$ is some
reference time scale such that, with $(t-s)/\tau_{\rm *}={\rm O}(1)$
\BD
\lim_{s\to \infty} R = \left\{
\begin{array}{ll}
R_{\rm eq}(t-s) & \mbox{\rm ~~;~ for $s/\tau_{\rm *}\to\infty$} \\
s^{-1-a} f_R(t/s) & \mbox{\rm ~~;~ for $s/\tau_{\rm *}={\rm O}(1)$}
\end{array}
\right.  
\ED 
Since in lattice calculations $s$ is always finite, the `crossover'
can be illustrated by studying $Q:= R(t,s)/R_{\rm eq}(t-s)\sim R(t,s)
(t-s)^{1+a}$. As long as LSI still describes the data, one expects
$Q\sim (y-1)^{a-a'}$ for $y=t/s\gtrapprox 1$ and deviations from it
should signal the presumed crossover to the quasistationary regime
$y\to 1$ where $Q(y)$ should become constant.  For the $1D$ critical contact
process we find that for $y=t/s\lessapprox 1.1$, $Q(y)$ still obeys
scaling for $s$ large enough and that $Q$ changes from $Q\approx 0.3$
at $y-1\approx 0.1$ to $Q\approx 0.8$ at $y-1\approx 10^{-4}$.  $Q(y)$
appears to become flatter as $y\to 1$, but the change to a
quasistationary behaviour could not yet be observed, in spite of large
waiting times $s>860000$, before strong finite-time effects set in at
$t-s={\rm O}(1)$.  In comparison, unpublished data for the $2D$ Ising
model \cite{Hinr06} show convergence towards $Q(y)\sim (y-1)^{0.187}$
as $s$ increases before finite-time effects destroy scaling.  We
conclude that LSI does accurately describe the data as long as $t/s$
is large enough such that the effects of the `crossover' are not yet
noticeable.  A quantitative analysis of data from the region
$t/s\lessapprox 1.1$ would require a precise theory of the `crossover'
between the ageing regime and the region $t-s\ll s$, the r\^ole of
finite-time effects and the influence of initial conditions, e.g.
different initial fillings of the lattice.

Very recently, a similar test was carried out in a $1D$ kinetic Ising model
with competing Glauber and Kawasaki dynamics \cite{Odor06}. The stationary
state is therefore not an equilibrium state. This was the first time that
LSI was tested and confirmed for a dynamics where the parity of the total
spin is conserved by the dynamics. 

Finally, we recall that studying the scaling behaviour of an
alternating susceptibility gives yet another direct access to the 
exponent $a'-a$. This was applied to 
the critical $3D$ Ising spin glass \cite{Henk05a}, 
with a binary distribution of the couplings $J_{i,j}=\pm J$. 

In summary, we have reconsidered the way how observables defined in 
non-equilibrium lattice models might be related to (quasi-)primary scaling
operators of field-theory. Our result eq.~(\ref{2}) points to a
so far overlooked subtlety which might be of relevance in the discussion of
the functional form of non-equilibrium scaling functions, for example in
ageing phenomena. It remains to be seen how general the phenomenon for which we have 
presented evidence really is.\footnote{It is not inconceivable that analogues might
exist in equilibrium critical phenomena, for instance when spatial
translation-invariance is broken by disorder or boundaries.} The results 
on $R(t,s)$ as collected in table~\ref{Tabelle1} for the non-equilibrium dynamics
of some models with $m_0=0$ appear to be 
compatible with the predictions eqs.~(\ref{R},\ref{Rq}) of local 
scale-invariance, provided cross-over effects to non-ageing 
regimes are negligible. The multitude of examples in table~\ref{Tabelle1}
suggests that rather being a kind of exotic exception (a belief
implicit in \cite{Henk02,Pico04,Henk05a}), the case  $a'\neq a$ 
might turn out to be the generic situation. Having seen that the
same mechanism also explains the exact autocorrelator of the $1D$
Glauber-Ising  model indicates that the correspondence (\ref{2}) should be
more than just a patching-up of data for the autoresponse function. 

What does this mean for the existence of local scale-invariance in
non-equilibrium dynamics, with $m_0=0$~? In a few exactly solved systems (where the
dynamical exponent $z=2$) we have found
exact agreement and in several models as generic as kinetic Ising models
or the contact  process eqs.~(\ref{R},\ref{Rq}) 
describe the data very well for $t/s$ not too small. It is 
remarkable that the two-time autocorrelations and autoresponses of 
models as physically different as those included in
table~\ref{Tabelle1} (and several further ones with $a=a'$ which we did not
include) can be described in terms of a single theoretical idea.
On the other hand, field-theoretical studies of the critical 
O($n$) model in both $4-\eps$ dimensions \cite{Cala05,Cala02} and in 
$2+\eps$ dimensions \cite{Fedo06}, although they agree with LSI
at the lowest orders in $\eps$, continue to find
discrepancies with either (\ref{R}) or (\ref{Rq}) at some higher order.
However, non-equilibrium field-theory presently only yields 
explicit results for the first few terms of the 
$\eps$-expansion series. When one truncates this series
to an $\eps$-dependent sum, the resulting numerical values for the scaling
functions are still far from the 
numerical data.\footnote{The second-order calculation in $4-\eps$ dimensions
for $n=1$ is a little closer to the numerical data than 
LSI with $a'=a$ \cite{Plei05}.}
But since we have shown that LSI reproduces the known exact results 
of both $R(t,s)$ and $C(t,s)$ of the $1D$ Ising model it might be too
simplistic to argue that LSI could at best describe gaussian fluctuations. 
A better understanding of the dynamical symmetries of non-equilibrium 
critical dynamics remains a challenging problem.  
  
\annexe{Two-time autocorrelations for $z=2$}

If the dynamical exponent $z=2$, local scale-invariance reduces to 
Schr\"odinger-invariance. We have already described in the past \cite{Pico04}
how two-time autocorrelation functions can be calculated in the case $\xi=0$
and we now wish to extend that treatment to the more general correspondence
(\ref{2}). We consider a Langevin equation of the form
$\partial_t \phi = -D\frac{\delta{\cal H}}{\delta \phi} - D v(t)\phi +\eta$
where $\cal H$ is the hamiltonian, $D$ the diffusion constant, the 
gaussian noise $\eta$ has  zero mean and variance 
$\langle \eta(t,\vec{r})\eta(s,\vec{r}')\rangle
=2DT\delta(t-s)\delta(\vec{r}-\vec{r}')$ and $T$ is the bath temperature. The
potential $v(t)$ acts as a Langrange multiplier which can be used to 
describe explicitly the breaking of time-translational invariance. Here we
restrict to situations where
\BEQ
k(t) := \exp\left[ -D  \int_0^t \!\D u\: v(u) \right] \sim t^{\digamma}
\EEQ
Then is has been shown \cite{Pico04} that for systems at criticality
\BEA
C(t,s) &=& \Bigl\langle \phi(t) \phi(s) \Bigr\rangle
= D T_c \int \!\D u\, \D\vec{R}\: 
\left\langle \phi(t,\vec{y})\phi(s,\vec{y}) \wit{\phi}^2(u,\vec{r}+\vec{y}) 
\right\rangle_0
\nonumber \\
&=& DT_c \int\!\D u\, \D\vec{R}\:
\frac{k(t)k(s)}{k(u)^2}{\cal R}_0^{(3)}(t,s,u;\vec{R})
\EEA
where ${\cal R}_0^{(3)}$ is the well-known three-point response function for
$v(t)=0$ which can be found from its Schr\"odinger-covariance and reads 
\cite{Henk94}
\BEA
\hspace{-1.5truecm}{\cal R}_0^{(3)}(t,s,u;\vec{r}) &\hspace{-0truecm}=& 
\hspace{-0.0truecm}{\cal R}_0^{(3)}(t,s,u) 
\exp\left[-\frac{\cal M}{2}\frac{t+s-2u}{(s-u)(t-u)}\, \vec{r}^2\right]
\Psi\left( \frac{t-s}{(t-u)(s-u)}\,\vec{r}^2\right)
\nonumber \\
\hspace{-1.5truecm} 
{\cal R}_0^{(3)}(t,s,u) &\hspace{-0truecm}=& 
\hspace{-0.0truecm}\Theta(t-u)\Theta(s-u) (t-u)^{-\wit{x}_2}
(s-u)^{-\wit{x}_2} (t-s)^{-x+\wit{x}_2}
\nonumber
\EEA
where $\Psi$ is an undetermined scaling function and the causality conditions
$t>u, s>u$ are noted explicitly. In writing this, we have dropped a term
coming from the correlations in the initial state which merely produces
finite-time corrections to the leading scaling behaviour, see
\cite{Bray94a,Cala05,Pico04}. 

We now generalize this to the primary scaling operators according to (\ref{2}). 
The operator $\Phi$ has the scaling dimension $x+2\xi$ and the composite
scaling operator $\wit{\Phi}^2$ has the scaling dimension
$2\wit{x}_2 + 4 \wit{\xi}_2$.\footnote{For bosonic free fields, one would have
$\wit{x}_2=\wit{x}$ and $\wit{\xi}_2=\wit{\xi}$.} We then obtain for the
physical autocorrelation function, up to normalization and with $t>s$
\BEA
\lefteqn{ 
C(t,s) = (ts)^{\xi} \int\!\D u\,\D\vec{R}\: 
\left\langle \Phi(t,\vec{y})\Phi(s,\vec{y})\wit{\Phi}^2(u,\vec{R}+\vec{y})
\right\rangle_0 u^{2\wit{\xi}_2}
}
\nonumber \\
&=& (ts)^{\xi} (t-s)^{-x-2\xi+\wit{x}_2+2\wit{\xi}_2-d/2} 
\int_0^s \!\D u\: \frac{k(t)k(s)}{k(u)^2} u^{2\wit{\xi}_2} 
\left[ (t-u)(s-u)\right]^{-\wit{x}_2-2\wit{\xi}_2+d/2} \nonumber \\
& & \times \int\!\D\vec{R}\: \exp\left[ -\frac{\cal M}{2} \frac{t+s-2u}{t-s}\, 
\vec{R}^2\right] \Psi\left(\vec{R}^2\right)
\nonumber \\
&=& s^{1+d/2-x-\wit{x}_2} \left(\frac{t}{s}\right)^{\xi+\digamma}
\left(\frac{t}{s}-1\right)^{\wit{x}_2+2\wit{\xi}_2-x-2\xi-d/2}
\nonumber \\
& & \times \int_0^1 \!\D v\, v^{2\wit{\xi}_2-\digamma} 
\left[\left(\frac{t}{s}-v\right)
\Big( 1-v \Big)\right]^{d/2-\wit{x}_2-2\wit{\xi}_2} 
\mathbf{\Psi}\left(\frac{t/s+1-2v}{t/s-1}\right)
\EEA
and where the function $\mathbf{\Psi}$ is defined by the 
integral over $\vec{R}$.
By comparison with the standard scaling from for  $C(t,s)$, we read off
$b=x+\wit{x}_2-1-d/2$ and $\lambda_C = 2(x-\digamma)+2\xi$.\footnote{A similar
calculation for the autoresponse function gives, up to normalization, 
\BD R(t,s) = s^{-(x+\wit{x})/2} (t/s)^{\xi+\digamma} (t/s-1)^{-x-2\xi}\,
\delta_{x+2\xi,\wit{x}+2\wit{\xi}}
\ED
which reproduces  again eq.~(\ref{R}), 
hence $\lambda_R=2(x-\digamma)+2\xi=\lambda_C$ as expected \cite{Bray94a} for
non-disordered systems without long-range initial correlations. In particular,
for critical systems with $a=b$ the equality $\lambda_C=\lambda_R$ implies that there
is a finite limit fluctuation-dissipation ratio 
$X_{\infty} = \lim_{(t/s)\to\infty} R(t,s)/(T_c \partial C(t,s)/\partial s)$, see
\cite{Cala05}.} 
Furthermore, since $1+a'=x+2\xi$, it turns
out that the form of the scaling function $f_C(y)$ is described  by just one
more parameter $\mu := \xi+\wit{\xi}_2$ and we finally have
\BEA
\lefteqn{
C(t,s) = C_0 s^b \left(\frac{t}{s}\right)^{1+a'-\lambda_C/2} 
\left(\frac{t}{s}-1\right)^{b-2a'-1+2\mu}  
}
\nonumber \\
& & \times \int_0^1 \!\D v\, v^{\lambda_C+2\mu-2-2a'} 
\left[ \left( \frac{t}{s}-v\right)\Big(1-v\Big)\right]^{a'-b-2\mu} 
\mathbf{\Psi}\left(\frac{t/s+1-2v}{t/s-1}\right)
\EEA
and we have also reintroduced a normalization constant $C_0$.
This should hold for simple (non-glassy) magnets with  $z=2$ and in situations
where the initial correlations have no effect on the leading scaling
behaviour; of course the scaling limit $s\to\infty$ and $t/s=y>1$ fixed is
understood. 

As an illustration, we consider the $1D$ Glauber-Ising model. At $T=0$, the
exact two-time autocorrelation function is \cite{Godr00}
\BEQ \label{gl:GIC}
C(t,s) = \frac{2}{\pi} \arctan\left( \sqrt{\frac{2}{t/s-1}} \right)
\EEQ
This holds true not only for the usually considered short-ranged
initial conditions but also for long-ranged initial spin-spin correlations
$\langle\sigma_r(0) \sigma_0(0)\rangle\sim r^{-\nu}$ 
with $\nu>0$ (for $\nu=0$ an analogous result holds
for the connected autocorrelator) \cite{Godr00}. In addition, the
exponents $a,a'$ and $\lambda_R$ are independent of $\nu$.  

In previous work \cite{Pico04}, we have already explained the form 
of the exact autoresponse function $R(t,s)$ in terms of the correspondence
eq.~(\ref{2}) (see table~\ref{Tabelle1}) but we had to leave open the 
analogous question for $C(t,s)$. In order to account for (\ref{gl:GIC}), 
we remark that for $t=s$, the autocorrelator should not be  singular. This
requires 
\BEQ \label{3}
\mathbf{\Psi}(w)=w^{b-2a'-1+2\mu} \;\; \mbox{\rm for $w\gg 1$}
\EEQ
The most simple way to realize this is to require that (\ref{3}) holds for 
all values of $w$. This kind of assumption was already seen to become exact
in models described by an underlying bosonic free field-theory \cite{Pico04}. 
Recalling from table~\ref{Tabelle1} 
that $b=a=0$ and $\lambda_C=1$ and assuming (\ref{3}) 
to hold for all $w$, we obtain
\BEQ \label{A7}
C(t,s) \approx  C_0 \int_0^1  \!\D v\, v^{2\mu} 
\left[ \left(\frac{t}{s}-v\right)\Big( 1-v\Big)\right]^{-2\mu-1/2} 
\left(\frac{t}{s}+1-2v\right)^{2\mu}
\EEQ
Because the exact result (\ref{gl:GIC}) is independent of  the initial
correlations, the comparison with the expression (\ref{A7}) derived from the
thermal noise is justified. 
The exact Glauber-Ising result
(\ref{gl:GIC}) is recovered from (\ref{A7}) 
for $\mu=-1/4$ and $C_0=\sqrt{2\,}/\pi$. 

This is the first example of an exactly solved  model with $a'\ne a$ 
where the scaling of
{\em both} the autoresponse and of the autocorrelation functions can be 
explained in terms of LSI. 

\noindent
{\bf Acknowledgements:} 
We thank J.L. Cardy, A. Gambassi, J.P. Garrahan, C. Godr\`eche, H. Hin\-rich\-sen, 
G.M. Sch\"utz and P. Sollich for discussions. 
M.H. thanks the Isaac Newton Institute and the INFN Firenze
for warm hospitality, where this work was done. T.E. is grateful for 
support by a Feoder Lynen fellowship of the
Alexander von Humboldt foundation and the Istituto Nazionale di Fisica
della Materia-SMC-CNR.
M.P. acknowledges the support by the Deutsche Forschungsgemeinschaft through
grant no. PL 323/2.
This work  was supported by the franco-german binational programme PROCOPE. 

  
\end{document}